\documentclass[prl,showpacs,superscriptaddress,lengthcheck,floatfix]{revtex4}

\usepackage{amsfonts}
\usepackage{amsmath}
\usepackage{amssymb}
\usepackage{graphicx}
\usepackage{graphics}

\begin{document}

\title{Predicted thermal superluminescence in low--pressure air}
\author{A.R. Aramyan}
\email{aramyan.artur@gmail.com}
\affiliation{Laboratory of Plasma Physics, Institute of Applied Problems of Physics,
Hrachya Nersessian str. 25, 375014 Yerevan, Armenia}
\author{K.P. Haroyan}
\affiliation{Laboratory of Plasma Physics, Institute of Applied Problems of Physics,
Hrachya Nersessian str. 25, 375014 Yerevan, Armenia}
\author{G.A. Galechyan}
\affiliation{Laboratory of Plasma Physics, Institute of Applied Problems of Physics,
Hrachya Nersessian str. 25, 375014 Yerevan, Armenia}
\author{N.R. Mangasaryan}
\affiliation{Department of physics, Russian--Armenian University, H. Emin str. 127, 375012 Yerevan, Armenia}
\author{H.B. Nersisyan}
\affiliation{Division of Theoretical Physics, Institute of Radiophysics and Electronics,
378410 Ashtarak, Armenia}
\date{\today }

\begin{abstract}
It is shown that due to the dissociation of the molecular oxygen it is possible to obtain
inverted population in low pressure air by heating. As a result of the quenching of the
corresponding levels of the atomic oxygen the thermal superluminescent radiation is generated.
It has been found that the threshold of the overpopulation is exceeded at the air temperature
$2300\lesssim T\lesssim 3000$ K. Using this effect a possible mechanism for the generation of
the flashes of the radiation in air observed on the airframe of the space shuttle during its
descent and reentry in the atmosphere is suggested.
\end{abstract}

\maketitle

In Ref.~\cite{ara03} the influence of the acoustic vortical motion on the spectrum of the optical
emission from the gas--discharge of argon at relatively high pressure has been experimentally studied.
The filling of the atomic levels of argon under influence of acoustic wave has been analyzed.
During the experiment the following picture has been observed. In several second after turning--off
the acoustic wave at the distinct sections of the discharge tube the flashes of light have been
observed during several minutes. It has been shown that the radiation is monochromatic and has
superluminescent origin. The reason of this effect is the quenching of the highly excited Rydberg
states which result in the inverted population of the atomic levels. In this experiment the Rydberg
atoms are formed due to the intense dissociative recombination of electrons and molecular ions
which is virtually the only process of bulk neutralization of charged particles in the gas--discharge
plasma at sufficiently high pressures ($p > 10$ mm Hg) \cite{ele82}. In this Letter we consider
the relatively low pressures ($\sim 1$ mm Hg) when the Rydberg atoms are formed by thermal dissociation
and propose a possible mechanism for generation of the superluminescent radiation by heating of air.
Based on this effect we show that the light flashes observed during the descent of the space shuttle
Columbia (see, e.g., \cite{int1}) are a result of the strong heating of air which induces thermal
superluminescence.

\begin{figure}[tbp]
\centerline{
\includegraphics[width=80mm,angle=90]{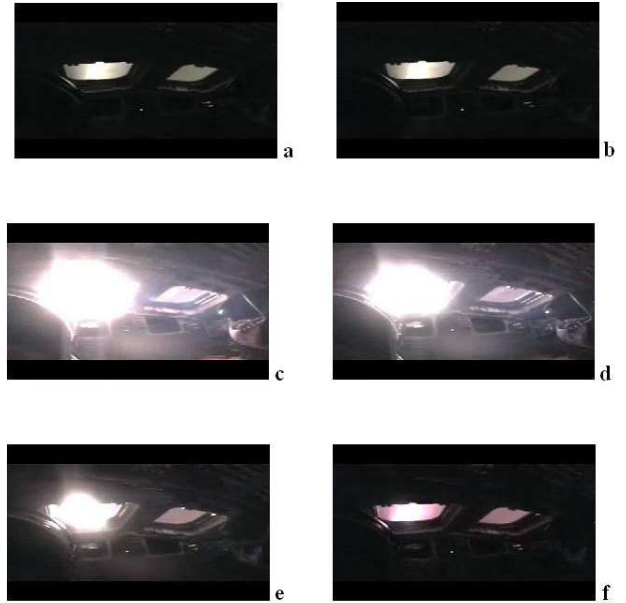}}
\caption{\label{f1}(Color online) Frame-by-frame evolution of the flash propagation. a) 0 ms, b) 40 ms,
c) 80 ms, d) 120 ms, e) 160 ms, f) 200 ms.}
\end{figure}
\begin{figure}[tbp]
\centerline{
\includegraphics[width=80mm]{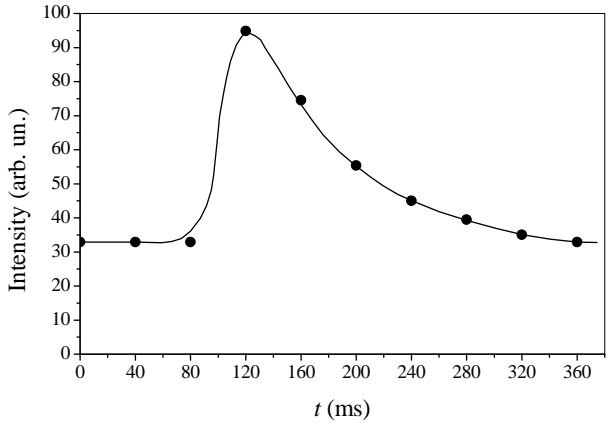}}
\caption{\label{f2}The radiation intensity of the flashes as a function of time.}
\end{figure}

It is known that during the descent of the space shuttle and its reentry in the atmosphere
the airframe is heated up to the temperatures $2000-3000$ K. Under such conditions the
flashes of the radiation are observed which is described in detail in Ref. \cite{int1}.
A frame-by-frame study of the videorecording of the flashes indicates (see Fig. \ref{f1})
that they are of orange color with the duration about $\sim 80$ ms and have an asymmetrical
temporal evolution, i.e. a steep front and a long tail (Fig. \ref{f2}). Such temporal
shape of the intensity indicates that the radiation is of stimulated nature.

For the formation of superluminescence it is imperative to invoke an overpopulation of the atomic levels
which are responsible for the radiation \cite{yar89}. Let us consider a possible mechanism for the
formation of a level overpopulation due to the heating of air. For this purpose one has to study
the possibility of the formation of highly excited Rydberg states of atoms in a heated air medium.

Consider a dissociative equilibrium in molecular (diatomic) gases i.e. an equilibrium
between atoms ($\mathrm{A}$) and molecules ($\mathrm{A}_{2}$) in the heated air which results from
the reaction
\begin{equation}
\mathrm{A}_{2} \Leftrightarrow 2\mathrm{A} .
\label{e1}
\end{equation}

It is known that the densities of the particles which are involved in the reaction \eqref{e1} obey
the equations which are similar to the Saha equation \cite{rai87,lan84}. The densities of atoms
$(n_{\mathrm{A}} )$ and molecules $(n_{\mathrm{A}_{2}} )$ which are formed and destructed due to the
dissociation, can be found using the mass action low \cite{rai87,lan84}
\begin{equation}
\frac{n^{2}_{\mathrm{A}}}{n_{\mathrm{A}_{2}}}=K_{\mathrm{A}} =\frac{g^{2}_{\mathrm{A}}}
{g_{\mathrm{A}_{2}}}
\frac{\omega _{\mathrm{A}_{2}}}{8J_{\mathrm{A}_{2}}}
\left(\frac{m^{3}_{\mathrm{A}}}{\pi^{3} \kappa T}\right)^{1/2} e^{-D_{\mathrm{A}_{2}}/\kappa T} ,
\label{e2}
\end{equation}
where $\kappa $ is the Boltzmann constant, $g_{\mathrm{A}}$ and $g_{\mathrm{A}_{2}}$ are the statistical weights
of the atom and molecule, respectively, $m_{\mathrm{A}} $ is the mass of the atom, $D_{\mathrm{A}_{2}}$ and
$J_{\mathrm{A}_{2}}$ are the dissociation energy and the moment of inertia of the molecule, respectively,
$\omega _{\mathrm{A}_{2}}$ is the frequency of its vibration, and $T$ is the gas temperature.

Equation \eqref{e2} should be supplemented with the condition of the conservation of the total
number of the particles
\begin{equation}
\frac{1}{2}n_{\mathrm{A}}+n_{\mathrm{A}_{2}} =n^{\ast}_{\mathrm{A}_{2}} ,
\label{e7}
\end{equation}
where $n^{\ast}_{\mathrm{A}_{2}}$ is the density of the initial molecular gas (before dissociation).
Solving the system of Eqs. \eqref{e2} and \eqref{e7} one obtains the density of atoms $n_{\mathrm{A}}(T)$
which are formed due to the dissociation of molecules,
\begin{equation}
n_{\mathrm{A}}(T) =\sqrt{\frac{1}{16}K_{\mathrm{A}}^{2}(T)+n^{\ast}_{\mathrm{A}_{2}} K_{\mathrm{A}}(T)}
-\frac{1}{4} K_{\mathrm{A}}(T) .
\label{e8}
\end{equation}
Here $K_{\mathrm{A}}(T)$ is given by Eq.~\eqref{e2}. Note that with increasing temperature up to
the value $\kappa T_{m}=2D_{\mathrm{A}_{2}}$ the density $n_{\mathrm{A}}(T)$ increases. However,
the preexponential factors in Eq.~\eqref{e2} are usually such, that the dissociation begins at the
temperatures much smaller than $T_{m}$. As a result all the molecules can be completely dissociated
at $T\ll T_{m}$.

On the basis of the relations \eqref{e2} and \eqref{e8} we consider now the dissociative equilibrium
in nitrogen and oxygen by the heating of air at the low pressure ($p\sim 1$ mm Hg). Using the values
of the parameters given in Table \ref{tab:1} (see, e.g., Refs. \cite{smi82,rai87,kik76}) we evaluate
these expressions for the atomic densities $n_{\mathrm{A}}(T)$ as the functions of air temperature.

\begin{table}[t]
\caption{\label{tab:1}The parameters of the nitrogen and oxygen molecules and atoms introduced in Eq. \eqref{e2}. Here
$B_{\mathrm{A}_{2}} =\hbar ^{2}/2J_{\mathrm{A}_{2}}$ is the rotational constant of the molecule.}
\begin{ruledtabular}
\begin{tabular}{llllll}
Molecule & $g_{\mathrm{A}}$  &  $g_{\mathrm{A}_{2}}$ & $D_{\mathrm{A}_{2}}$ (eV) & $\hbar \omega_{\mathrm{A}_{2}}$ (eV) &
$B_{\mathrm{A}_{2}}$ (meV) \\
\hline
N$_{2}$ & 4 & 1 & 9.76 & 0.2926 & 0.2492  \\
O$_{2}$ & 9 & 3 & 5.08 & 0.2326 & 0.2073  \\
\end{tabular}
\end{ruledtabular}
\end{table}

\begin{figure}[tbp]
\centerline{
\includegraphics[width=80mm]{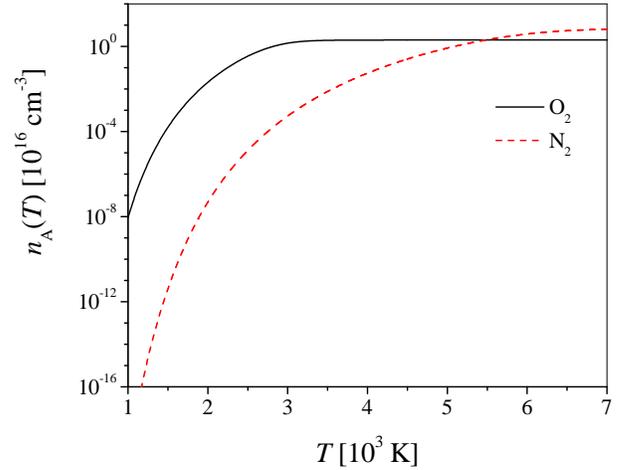}}
\caption{\label{f3}(Color online) The densities of the oxygen ($n _{\mathrm{O}}(T)$, solid lines) and nitrogen ($n _{\mathrm{N}}(T)$,
dashed line) atoms as a function of air temperature.}
\end{figure}

Figure~\ref{f3} shows the densities  $n_{\mathrm{O}}(T)$ and $n_{\mathrm{N}}(T)$ (in units
$10^{16}$ cm$^{-3}$) as the functions of temperature (in units $10^{3}$ K) assuming that
$n^{*}_{\mathrm{N}_{2}}\simeq 3.54 n^{*}_{\mathrm{O}_{2}}$. Also it is assumed that the
relative quantities of oxygen ($\sim 22$ percent) and nitrogen ($\sim 78$ percent) in air
are independent of the altitude. The density of the initial molecular oxygen is
$n^{*}_{\mathrm{O}_{2}}=10^{16}$ cm$^{-3}$ which at the temperatures $1000<T<4000$ K corresponds
to the partial pressures $1\lesssim p\lesssim 4$ mm Hg of the gas. It is seen that in the temperature domain
$1000 \leqslant T\leqslant 4000$ K the density of the atomic oxygen $n_{\mathrm{O}}(T)$ much larger
than the density of the atomic nitrogen $n_{\mathrm{N}}(T)$. Consequently for a given temperature the
dissociation rate for the nitrogen in air is much smaller compared to the atomic oxygen. Therefore
the formation of the highly excited states we shall further study only for oxygen.

We propose that the orange flashes of the radiation observed during the descent of the space shuttle and its
reentry in the atmosphere are indeed the consequence of the transitions in oxygen atoms while the contributions
of the nitrogen atoms are negligible. The transitions in the oxygen atom which yield orange radiation correspond
to the $5d\to 3p$ and $6s\to 3p$ transitions.

The early experiments (see Refs.~\cite{fre71,kup69}) showed that in the discharge chamber about 3--5 percent
of the atoms formed due to the dissociation of the molecular oxygen are in the highly excited Rydberg
states. This process was widely used in early investigations of Rydberg atoms.

Let us consider the efficiency of the filling of the levels for the $5d \to 3p$ and $6s \to 3p$ transitions
in atomic oxygen versus gas temperature. In other word we study the probability of the thermal overpopulation
of the levels $5d$ and $6s$.

High quantum levels ($n\gg 1$) of the excited atoms are known from the quantum theory to have large lifetimes.
The average lifetime $t_{n}$ of the excited level depends on the principal quantum number $n$ as $t_{n} \sim n^{3}$
\cite{bet57}.

Consider now the destruction of such highly excited long--lived atomic states as a result of collisions with
atoms and molecules.

According to the theory of atomic collisions \cite{mot65,smi82}, the transition probability between two
atomic states due to the collisions is given by $w_{nl}\simeq (C/v^{2}_{a}) \exp (-\xi )$ and strongly depends
on the parameter $\xi $. Here $v_{a}$ is the relative velocity of the nuclei of the colliding atoms, $C$
is some constant. Let us estimate the parameter $\xi $ for the transition $nl \to n'l'$, where $n'=n-1$.
The energy of this transition at $n\gg 1$ is given by
\begin{equation}
\Delta\varepsilon \sim \frac{1}{n^{3}} \left(\delta_{l}-\delta_{l'}\right) ,
\label{e5}
\end{equation}
where $\delta_{l}$ is the quantum defect. The parameter $\xi $ is defined as \cite{mot65,smi82}
\begin{equation}
\xi =\Delta\varepsilon \frac{a_{n}}{\hbar v_{a}}\sim \frac{v_{B}}{v_{a}} \frac{\delta_{l}}{n} .
\label{e6}
\end{equation}
Here $a_{n}\sim n^{2}$ is the radius of the excited atom, $v_{B} $ is the Bohr's velocity. In the case
of $n,l\gg 1$ the parameter $\xi $ turns out small due to the smallness of the quantum defect $\delta_{l} $ and
the probability $w_{nl} $ of the corresponding transitions increases (at the fixed temperature,
$v_{a}\sim T^{1/2}$). The situation is different when the highly excited levels have a small orbital
quantum number $l\geqslant 0$. In these cases, the parameter $\xi $ increases for the moderate values of
$n$. Then the probability of the transitions due to the collisions is much smaller than that in the
former case. It should be emphasized that in the temperature domain which is considered here
($\kappa T\ll D_{\mathrm{O}_{2}} $) the transition probability $w_{nl} $ increases with $T$, which
intensifies the processes of the transitions from the upper levels to the lower ones. Thus owing to
the quenching of highly excited states due to the collisions with atoms and molecules, the levels
with $n,l\gg 1$ are quickly emptied, whereas the levels with $n >1$, $l\geqslant 0$ are quickly occupied.
In our case, the levels $6s$ and $5d$ are occupied. Concerning the flashes and their spatial location
of the formation the observed superluminescent radiation is formed when the threshold of the overpopulation
is exceeded \cite{yar89}.

\begin{figure}[tbp]
\centerline{
\includegraphics[width=80mm]{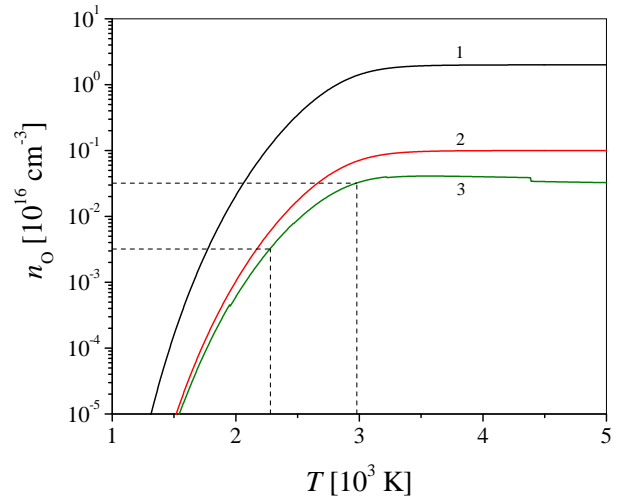}}
\caption{\label{f4}(Color online) The density of the dissociated oxygen atoms as a function of air temperature at
$n^{*}_{\mathrm{O}_{2}}=10^{16}$ cm$^{-3}$ (line 1). The line 2 corresponds to the density of the highly
excited Rydberg atoms, i.e. $n^{*}_{\mathrm{O}}(T)=0.05n_{\mathrm{O}}(T)$. The density of the Rydberg
atoms taking into account the thermal depletion of the energy levels
is shown as a line 3.}
\end{figure}

It is possible to estimate the overpopulation of the levels $5d$ and $6s$ with respect to $3p$. For
this purpose we employ Eqs.~\eqref{e2} and \eqref{e8} which define the density of the dissociated
oxygen atoms as a function of air temperature. The density $n_{\mathrm{O}}(T)$ is shown in Fig.~\ref{f4}
(line 1). [Note that this curve is the same as the solid line in Fig.~\ref{f3}]. From this figure it is
seen that the density of the atoms becomes appreciable at $T>2000$ K and at higher temperatures ($T>3000$ K)
the molecules of oxygen are practically dissociated. In addition, the line 2 demonstrates the density
$n^{*}_{\mathrm{O}}(T)=0.05n_{\mathrm{O}}(T)$, i.e. five percent of the density of the dissociated atoms
which, according to the experiments \cite{fre71,kup69}, are in highly excited Rydberg states.

\begin{figure}[tbp]
\centerline{
\includegraphics[width=75mm]{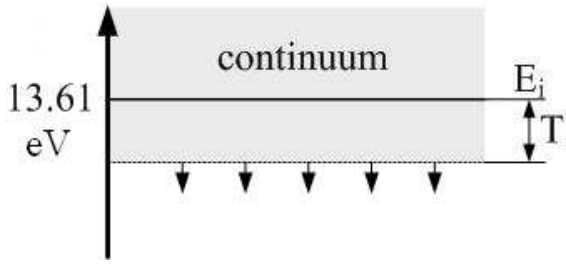}}
\caption{\label{f5}The energetic interval $E_{i}-\kappa T<E<E_{i}$, where the atomic energy levels are depleted.}
\end{figure}

As has been mentioned above the density of the highly excited atoms decreases with increasing temperature
due to the collisions with atoms and molecules. Moreover, the probability of the thermal ionization of
such atoms is quite high and with increasing temperature the highly excited levels are emptied.
Consequently the density $n_{\mathrm{Ry}}(T)$ of the Rydberg atoms can essentially differ
from $n^{*}_{\mathrm{O}}(T)$. For the qualitative analysis of this effect it is assumed here that in
the interval $E_{i}-\kappa T <E< E_{i}$ of the discrete energies, where $E_{i}\simeq 13.614$ eV is
the ionization energy of the atomic oxygen \cite{kik76}, the atomic levels become empty (see Fig.
\ref{f5}). Then using the Boltzmann distribution the total density of the Rydberg atoms which are
on the level $5d$ and higher can be approximated by
\begin{equation}
n_{\mathrm{Ry}}(T)=n^{*}_{\mathrm{O}}(T)\frac{\sum ^{\prime}_{n}g_{n}e^ {-E_{n}/\kappa T}}
{\sum_{n} g_{n}e^{-E_{n}/\kappa T}} .
\label{eq:ryd}
\end{equation}
Here $g_{n}$ and $E_{n}$ are the statistical weight and the energy of the atomic level $n$, respectively.
This quantities can be found, for instance, in the online atomic data database in Ref.~\cite{int2}.
In the denominator of Eq.~\eqref{eq:ryd} the summation is performed over all atomic levels starting with
the ground level whereas in the numerator the sum is performed starting with the level $5d$ and is
restricted by $n\leqslant n_{\max}(T) \simeq (E_{i}/\kappa T)^{1/2}$ according to the assumption
mentioned above. However, in the temperature domain considered here the contribution of the levels
below $5d$ to the statistical sum in the numerator of Eq.~\eqref{eq:ryd} provided to be negligible
(see below). Therefore the summation over $n$ in the numerator of Eq.~\eqref{eq:ryd} can also be
started at the level $5d$. The density of the highly excited atoms $n_{\mathrm{Ry}}(T)$ calculated
with the help of Eq.~\eqref{eq:ryd} is shown in Fig.~\ref{f4} (line 3) as a function of the air
temperature. From this figure it follows that with increasing of the air temperature up to $\sim 3000$ K
the density of the Rydberg atoms rapidly increases. The further increase of the temperature results
in the decrease of the density of the highly excited atoms. Thus the density of the Rydberg atoms
has maximum at the temperature $\sim 3000$ K.

In addition taking into account the possible mechanism suggested in Ref.~\cite{ara03} for the
quenching of the Rydberg states in the case of superluminescent flashes in the argon discharge
one can conclude that the highly excited atoms gather basically on the levels $6s$ and $5d$.

Consider the possibility of the overpopulation of the levels $6s$ and $5d$ relative to $3p$
in the atomic oxygen. It has been obtained above that the levels $6s$ and $5d$ are occupied
due to the quenching of the Rydberg atoms. Population of the level $3p$ (relative to the ground
state) can be estimated employing the Boltzmann distribution. Using the statistical weights
for the excited ($g_{3p}=3$) and the ground ($g_{0}=1$) states, as well as the energy of the
level $3p$ with respect to the ground state, $E_{3p}\simeq 10.85$ eV (see, e.g., Ref. \cite{int2}),
one obtains that the occupation of $3p$ is practically negligible up to the temperatures $3500$ K.
Thus after the dissociation of the molecules of oxygen the overpopulation of the levels $6s$ and
$5d$ relative to $3p$ is formed.

We calculate now the threshold of the overpopulation which leads to the generation of the
superluminescent radiation and the corresponding critical temperature. For this purpose we
employ a formula for the threshold value of the density for the population inversion obtained
in Ref. \cite{yar89},
\begin{equation}
\Delta n_{c} =\frac{8\pi\mathsf{n}^{3}\nu ^{2} t_{sp} \Delta\nu}{c^{3}t_{c}} ,
\label{e11}
\end{equation}
where $\mathsf{n}$ is the refractive index of the gas in the frequency range far from the resonance,
$\nu $ and $\Delta\nu =1/2\pi t_{c}$ are the frequency and the line width of the radiation, respectively,
$t_{c}$ is the photon lifetime, $t_{sp}$ is the spontaneous lifetime. Taking into account that in the
present case the spontaneous time $t_{sp}$ varies in the interval $10-100$ $\mu$s (see Ref. \cite{fre71}),
from Eq. \eqref{e11} we obtain $3.2\times 10^{13}\lesssim \Delta n_{c} \lesssim 3.2\times 10^{14}$ cm$^{-3}$.
According to Fig. \ref{f4} (line 3) such densities are obtained at the temperatures of air $2300\lesssim %
T\lesssim 3000$ K (see the dashed lines in Fig. \ref{f4}) which are in good agreement with the expected range
of the temperatures of the airframe during the reentry of the space shuttles in the atmosphere.

In conclusion, we have shown that it is possible to obtain the highly excited Rydberg atomic states by
heating of air which are formed due to the dissociation of the molecular oxygen. The quenching of these
states due to the collisions with atoms and molecules may result in the formation of the overpopulation;
The subsequent excess of the threshold density of the inversion leads to the generation of the flashes
of superluminescent radiation.
Furthermore, based on this result we have demonstrated that the flashes of the radiation
observed during the descent of the space shuttles and their reentry in the atmosphere have superluminescent
origin which are formed in heated air on the airframe. Finally, we believe that the effect considered here
can be realized in laboratory experiments. This will provide a useful benchmark on testing the present
qualitative treatment as well as for understanding some similar atmospheric phenomena. In addition,
this will enable us to vary in a wide range the parameters of the experiment (such as the density of
the gas) and investigate in more detail the intensity and the spectrum of the superluminescent radiation.

The authors are grateful to N.V. Tabiryan for useful discussions on the current problem.
Also it is a pleasure to acknowledge the help from Amal K. Das for preparation of the
manuscript.

\end{document}